\documentclass[prl,amsmath,amssymb,aps,twocolumn]{revtex4-1}
\usepackage{graphicx}
\usepackage{color}
\usepackage{amsmath}
\usepackage{mathrsfs}
\usepackage{amssymb}
\usepackage{amsthm}
\usepackage{physics}
\usepackage{notes2bib}
\usepackage{booktabs}
\usepackage[dvipdfm,colorlinks,urlcolor=blue,linkcolor=blue,anchorcolor=blue,citecolor=blue]{hyperref}
\usepackage{lineno}

\begin{document}


\title{The effect of dressing on thermalization of interacting waves}

\author{Zhen Wang}
\author{Yong Zhang}
\author{Hong Zhao}\email{zhaoh@xmu.edu.cn}

\affiliation{Department of Physics, Xiamen University, Xiamen 361005, Fujian, China}

\date{\today }
\begin{abstract}
We propose a more general setup for prethermalization in the system of interacting waves. The idea lies in dividing the multi-wave interactions into trivial and nontrivial ones. The trivial interactions will dress waves and lead to a less strongly interacting system which is statistically equivalent to the
original one. With this in mind, we find that prethermalization occurs not only in the weakly interacting regime but also in the strongly interacting regime. The irreversible process towards equilibrium is governed by the Zakharov equation, from which double scaling of the thermalization time is expected. Finally, the theory is well confirmed in numerical experiments.
\end{abstract}

\maketitle
\emph{Introduction}.---For a many-body system, a highly atypical state evolving under Hamiltonian dynamics is generally expected to thermalize \cite{DAlessio2016,Gogolin2016,Mori2018,Fu2019,Pistone2019}. Extensive and intensive research has shown that \emph{prethermalization} is a typical feature for a very general class of weakly interacting systems and that the time needed to reach thermalization is inversely proportional to the squared perturbation strength \cite{Reimann2019,Mallayya2019,Mallayya2021,Bastianello2020,Fu2019,Pistone2019,Huveneers2020}. However, relatively scant theoretical research studies the strongly interacting systems.  This work tackles the issue in classical systems and shows that prethermalization survives the strong interaction.

The Hamiltonian of interest has the form ${H} =H_0+g V$, where $H_0$ is usually an integrable Hamiltonian and $g V$ breaks integrability.  How such a system reaches a thermal state is an intriguing and fundamental question in physics. For classical systems, the study of this issue can be traced back to the seminal work of Fermi, Pasta, Ulam, and Tsingou (FPUT) in the 1950s \cite{Fermi1955}. Since then, continued efforts have been made to address this issue \cite{Ford1992,Lepri1998,Gallavotti2008,Benettin2008,Benettin2011,Benettin2013}.  Until recently, theoretical studies figure out that multi-wave resonances allow for energy redistribution and thus lead to thermalization. This irreversible process is described by the Zakharov equation, from which it is found that the system reaches thermalization on a timescale $\propto {1}/{g^2}$ \cite{Onorato2015,Pistone2018,Lvov2018,Fu2019,Pistone2019,Wang2020}.  The same results have been reported for a general class of quantum systems, but at this time, the Zakharov equation is superseded by its quantum version---the quantum Boltzmann equation \cite{Mori2018,Mallayya2019,Mallayya2021}.

The success in theory depends on the prerequisite that the perturbation is weak, i.e., $g\ll 1$. Such a perturbation ensures that the system quickly settles to a quasi-stationary state encoded via a generalized Gibbs ensemble (GGE) \cite{Yuzbashyan2016} and relaxes to the final thermal state on a much longer timescale $\propto {1}/{g^2}$. This time separation is termed prethermalization . When it comes to the strong perturbation, Lepri found an estimate for the timescale needed to reach thermalization \cite{Lepri1998}. Nevertheless, unlike in the case of weak perturbation, the underlying mechanism remains largely unexplored. What role dose multi-wave resonances play now? Moreover, what are the relevant timescales? It is questionable whether prethermalization survives.

In this paper, we address these questions in a common setup dealing with interacting waves. By dividing the multi-wave interactions into trivial and nontrivial ones, we map the system of interacting waves into a system of interacting dressed waves. The latter is statistically equivalent to the former but with weaker interactions. Taking FPUT-like models as examples,  we show that prethermalization survives the strong interaction and that thermalization still is dominated by resonances among waves, now merely dressed. The Zakharov equation describing the effect of resonances predicts the double scaling of the thermalization time, which is verified by numerical experiments.

\emph{Setup}.---In what follows, we will use a vector $\boldsymbol{a}$ to denote a set of variables $\{a_1,a_2,\ldots,a_1^*,a_2^*,\ldots\}$. We consider a generic setup dealing with interacting waves (quasi-particles).  The Hamiltonian reads
\begin{equation}\label{eq:original Hamiltonian}
{H} = H_0(\vb*{a})+g V^{(p)}(\vb*{a}),\; H_0(\vb*{a})=\sum_{k}\omega_ka_k^{*}a_k,
\end{equation}
where $\omega_k$ is the frequency spectrum, and $k$ denotes momentum modes. The term $H_0$ is the so-called unperturbed Hamiltonian and usually integrable. For the integrability-breaking term, we adopt the usual $p$-wave interactions, i.e.,
\begin{equation}\label{eq:integrability breaking term}
  V^{(p)}(\vb*{a})=\dfrac{1}{p!} \sum_{ \vb*{j},\vb*{l}\atop q+r=p} V_{\vb*{j}}^{\vb*{l}}a_{j_1} \cdots a_{j_q} a_{l_1} ^*\cdots a_{l_r} ^*+\mathrm{c.c.},
\end{equation}
whose strength is modulated by $g$. Here and below, shorthand notations are used: $a_{j}=a_{k_{j}}$; $\vb*{j}=(j_1,\cdots,j_q)$, $\vb*{l}=(l_1,\cdots,l_r)$; and thus $V_{\vb*{j}}^{\vb*{l}}=V_{j_{1}  \ldots j_{q} } ^{l_{1}  \ldots l_{r} }$. For translation-invariant interactions, $V_{\vb*{j}}^{\vb*{l}}$ is nonzero only when $k_{j_1}+\cdots+k_{j_q}=k_{l_1}+\cdots+k_{l_r}$. In analogy with quantum mechanics, we can view $\vb*{a}$ as \emph{incoming} waves and $\vb*{a}^*$ as \emph{outgoing} waves. Then, the term  $a_{j_1} \cdots a_{j_q} a_{l_1} ^*\cdots a_{l_r} ^*$ can be interpreted as a process in which $r$ outgoing waves are ``created'' as a result of interacting among $q$ incoming waves.

Suppose the dynamics is initiated from a state $f_{\mathrm{I}}$. The system is isolated during $t>0$ so that the phase-space distribution evolves as
$f(t)=e^{-Lt}f_{\mathrm{I}}$, where $L$ is the Liouville operator, defined in terms of the Poisson bracket as $L=\pb{{H}}{\;}$.  After a sufficiently long-time evolution, the system is expected to equilibrate at its thermal state given by a Gibbs ensemble (GE)\cite{Monacelli2021}
\begin{equation}\label{eq:Gibbs ensemble}
  f_{\mathrm{GE}}=\frac{e^{-{H}/\theta}}{\Trace[e^{-{H}/\theta}]},
\end{equation}
where $\Trace[f]$ is the normalizing partition function, and $\theta$ is the equilibrium temperature.

To describe the expectation values of observables after equilibration when $H$ is nonintegrable, one usually replace the GE \eqref{eq:Gibbs ensemble} with a most general Gaussian distribution \cite{Monacelli2021}
\begin{equation} \label{eq:202202072114}
  \tilde{f}_{\mathrm{eq}}\qty(\vb*{a}) =\frac{e^{-\tfrac{1}{2} (\vb*{a} ^\dagger-\vb*{\alpha} ^\dagger)\cdot\Theta^{-1}\cdot
  (\vb*{a} -\vb*{\alpha} )} }{\Trace[e^{-\tfrac{1}{2} (\vb*{a} ^\dagger-\vb*{\alpha} ^\dagger)\cdot\Theta^{-1}\cdot
  (\vb*{a} -\vb*{\alpha} )} ]} ,
\end{equation}
with mean vector $\vb*{\alpha} =\expval*{\vb*{a} } _{\mathrm{GE} }$ and covariance matrix
$\Theta =\expval*{\qty(\vb*{a} ^{\dagger} -\vb*{\alpha} ^{\dagger} )\qty(\vb*{a} -\vb*{\alpha} )} _{\mathrm{GE} }$. This is a result of the mean-field theory. The subscript `GE' here indicates that averaging is performed under GE \eqref{eq:Gibbs ensemble}. Without losing generality, in what follows, we will restrict the discussion to cases where $\vb*{\alpha}=0$.

The hermiticity of matrix $\Theta$ tells that there exists a unitary matrix $U$ such that $[U ^ {\dagger} \Theta U]_{kl} = \theta \delta_{kl}/\tilde{\omega}_k$. It is rewarding to introduce a set of new coordinates $\tilde {\vb*{a}} = U \vb*{a}$, in terms of which the distribution \eqref {eq:202202072114} is rewritten as
\begin{equation} \label{eq:202202131615}
  \tilde{f} _{\mathrm{eq} }(\tilde{\vb*{a}})
=\frac{e^{-\sum_{k} \tilde{\omega} _k\tilde{a} _k^{*}\tilde{a} _k/\theta} }{\Trace[e^{-\sum_{k} \tilde{\omega} _k\tilde{a} _k^{*}\tilde{a} _k/\theta}]}.
\end{equation}
This distribution is the stationary state of an integrable Hamiltonian system
\begin{equation} \label{eq:202202131624}
  \tilde{H}_0\qty(\vb*{a})=\sum_{k} \tilde{\omega} _k\tilde{a} _k^*\tilde{a} _k,
\end{equation}
which describes a system of noninteracting dressed waves.

The general equipartition theorem \cite{Beale2011} relates that $\expval*{a_k{\partial H}/{\partial a_l}}_{\mathrm{GE}}=\theta\delta_{kl}$ and $\expval*{a_k{\partial H}/{\partial a_l^*}}_{\mathrm{GE}}=0$, from which we obtain
\begin{subequations}
\begin{align}\label{eq:correlation}
 \expval{a_ka_l^*}_{\mathrm{GE}} &= \frac{\theta}{\omega_k}\delta_{kl}-\frac{g}{\omega_k}\expval*{a_k{\partial V^{(p)} \qty(\vb*{a} )}/{\partial a_l} }_{\mathrm{GGE}},  \\
  \expval{a_ka_l}_{\mathrm{GE}}  &=-\frac{g}{\omega_k}\expval*{a_k{ \partial V^{(p)}(\vb*{a})}/{\partial a_l^*}}_{\mathrm{GGE}},
\end{align}
\end{subequations}
where we replace GE with GGE when averaging on the right side. For a system of noninteracting waves, i.e., $g=0$, the waves remains free, and hence matrix $\Theta$ itself is diagonal. However, when the nonlinear interaction enters, waves become correlated, and, in general, $\Theta$ is no longer diagonal. Usually not all interactions contribute to the covariance. For this reason, we may divide the interactions $V^{(p)}(\vb*{a})$ into two groups, i.e.,
\begin{equation}\label{eq:split perturbation}
  V^{(p)}(\vb*{a})=V_{0}^{(p)}\qty(\vb*{a})+V_{\mathrm{non}}^{(p)}\qty(\vb*{a}),
\end{equation}
where $V_{0}^{(p)}(\vb*{a})$, embracing all trivial interactions, contributes to the covariance and $V_{\mathrm{non}}^{(p)}(\vb*{a})$ does not.

The above analyses show that the statistical information embedded in $H_0$ and $gV_{0}^{(p)}$ is encoded into $\tilde{H}_0$. Reorganizing the original Hamiltonian \eqref{eq:original Hamiltonian}, we will acquire a Hamiltonian that is statistically equivalent to it. The result is
\begin{equation} \label{eq:202202131626}
  \tilde{H}\qty(\tilde{\vb*{a}}) =\tilde{H} _0\qty(\tilde{\vb*{a}})+ g \tilde{V}^{(p)}_{\mathrm{non}}\qty(\tilde{\vb*{a}}).
\end{equation}
Suppose $g\tilde{V}^{(p)}\ll H_0$, which is usually true, we map the original system \eqref{eq:original Hamiltonian} into a more weakly interacting one. For such a system \eqref{eq:202202131626}, at the large timescale ($t\gg 2\pi/\tilde{\omega}_k$), the phase-space distribution $\tilde{f}(t)$ can be approximated by a GGE
\begin{equation} \label{eq:202202131615}
  \tilde{f} _{\mathrm{GGE} } \qty(t)
=\frac{e^{-\sum_{k} \tilde{\omega} _k\tilde{a} _k^{*}\tilde{a} _k/\theta_k(t)} }{\Trace[e^{-\sum_{k} \tilde{\omega} _k\tilde{a} _k^{*}\tilde{a} _k/\theta_k(t)}]}.
\end{equation}
If the system thermalizes, one expects that $\theta_k(t)=\theta$ and the distribution \eqref{eq:202202131615} results. Due to  statistical equivalence, we expect that the GGE \eqref{eq:202202131615} can replace the phase-space distribution $f(t)$ of the original system. This result is independent of the interaction strength and is an extension of the ansatz
\begin{equation} \label{eq:202202131615}
 {f} _{\mathrm{GGE} } \qty(t)
=\frac{e^{-\sum_{k} {\omega} _k{a} _k^{*}{a} _k/\theta_k(t)} }{\Trace[e^{-\sum_{k} {\omega} _k{a} _k^{*}{a} _k/\theta_k(t)}]},
\end{equation}
which only makes sense for weak interactions.

The map $H\rightarrow \tilde{H}$ provides an alternative path to identify dynamics of observables $O$. Next we will give the second-order approximation to $\expval*{\dot{O}}$. To this end, we now split the Liouville operator of the equivalent system \eqref{eq:202202131626}. The result reads
\begin{equation}\label{split of Liouville operator}
  \tilde{L}=\tilde{L}_0+\tilde{L}_V, \hbox{ with } \tilde{L}_0=\pb*{\tilde{H}_0}{\;}, \tilde{L}_V= g\pb*{\tilde{V}}{\;}.
\end{equation}
Making some necessary approximations \cite{Zwanzig2001}, we obtain
\begin{equation} \label{eq:202202072138}
\begin{aligned}
  \pdv{ \tilde{f} }{ t} \approx -\tilde{L} _0\tilde{f}(t)&-\tilde{L} _Ve^{-\tilde{L} _0t} \tilde{f} _0\\
  &+\tilde{L} _V\int_{0} ^{t} \tilde{L} _V\qty(t-s)\tilde{f}(t)\dd{s},
\end{aligned}
\end{equation}
where $\tilde{L} _V(t)=e^{-\tilde{L} _0t} \tilde{L} _V e^{\tilde{L} _0t} $. We restrict our attention to only those situations where the initial distribution is $\tilde{f}_{\mathrm{I}}$ such that ${\partial\tilde{f}_{\mathrm{I}}}/{\partial t}=\tilde{L}_0\tilde{f}_{\mathrm{I}}=0$. A straightforward calculation gives the time-dependent average of observable $O$. The result is
\begin{equation}\label{eq:202202072136}
\begin{aligned}
  \expval*{ \dot{O}} \approx \expval*{ \tilde{L}_0 O} _{\mathrm{GGE}} &+\expval*{ \tilde{L} _VO} _{\tilde{f}_0}\\
  &+ \int_{0} ^{t} \expval*{ \tilde{L} _V\tilde{L} _V\qty(t-s)O} _{\mathrm{GGE}} \dd{s} .
  \end{aligned}
\end{equation}
where $\tilde{f}(t)$ has been replaced with the GGE \eqref{eq:202202131615} when averaging.

\emph{Models and observables}.---
To facilitate the discussion and the comparison with established numerical experiments, we use the FPUT-like model as example. The Hamiltonian is
\begin{equation}\label{eq:reference Hamiltonian}
 \mathscr{H}_p=\sum_{j=1}^{N}\dfrac{\dot{u}_j^2}{2}+\dfrac{1}{2}\left(u_{j+1}-u_{j}\right)^2+\dfrac{\lambda}{p!}\qty(u_{j+1}-u_{j})^p,
\end{equation}
where $\dot{u}_j$ and $u_j$, respectively, denote the velocity and the displacement of particle $j$ from the equilibrium position. In Eq.~\eqref{eq:reference Hamiltonian}, the sum of first two terms constitutes our unperturbed Hamiltonian $H_0$, and that of the nonlinear terms breaks the integrability. Stability in strongly interacting regime imposes that $p$ is even.

After some transformations \cite{Wang2021}, the Hamiltonian has the general form \eqref{eq:original Hamiltonian}. The interaction strength now is closely related to energy density $\varepsilon=E/N$ and nonlinearity strength $\lambda$, specifically, $g=\lambda\varepsilon^{(p-2)/2}$. For a FPUT-like model, when the nonlinearity is present, the waves $a_k$ and $a_{N-k}$ become correlated, i.e., $\langle a_ka_{N-k}\rangle \ne 0$ \cite{Gershgorin2007,Wang2021}. The unitary transformation which diagonalizes the matrix $\Theta$ has been given by
\begin{equation}\label{eq:transformation}
  a_k=\left(\sqrt{\eta}+\dfrac{1}{\sqrt{\eta}}\right)\tilde{a}_k+\left(\sqrt{\eta}-\dfrac{1}{\sqrt{\eta}}\right)\tilde{a}^*_{N-k},
\end{equation}
where $\eta$ is called renormalization factor. The asymptotic behavior of $\eta$ is given in the SM. The statistically equivalent Hamiltonian now reads
\begin{equation}\label{effective Hamiltonian}
  \tilde{\mathscr{H}}_p=\sum_{k}\tilde{\omega}_k\tilde{a}_k^*\tilde{a}_k+\frac{g}{\eta^{p/2}}{V}^{(p)}_{\mathrm{non}}(\tilde{\boldsymbol{a}}),
\end{equation}
where $\tilde{\omega}_k=\eta\omega_k$. The interaction ${V}^{(p)}_{\mathrm{non}}$ excludes all the trivial interactions, effectively weakening the nonlinear interactions ($\eta\ge 1$) \cite{Gershgorin2007,Lee2009,Gershgorin2005}.

The Eq.~\eqref{eq:202202131615} tells that $\expval*{\tilde{a}_k^*\tilde{a}_l}_{\mathrm{GE}}=\theta/\tilde{\omega}_k\delta_{kl}$ when the system thermalizes. For this reason, we study the dynamics of the wave action spectral density $\tilde{n}_k=\expval*{\tilde{a}_k^*\tilde{a}_k}$. Substituting $O=\tilde{a}_k^*\tilde{a}_k$ into Eq.~\eqref{eq:202202072136} gives
\begin{equation}\label{eq:n-wave kinetic equation}
\dot{\tilde{n}}_k=\eta_{k}-\gamma_k \tilde{n}_k.
\end{equation}
Here we need to emphasize that this result is self-consistent only when the linear and nonlinear timescales are separated, i.e., $2\pi/\tilde{\omega}_k\ll 1/\gamma_k$. Explicit expressions for coefficients $\eta_{k}$ and $\gamma_k$,  and details of our derivation are
given in the SM \cite{Wang2021}. As we demonstrate in the SM, a general H-theorem shows that Eq.~\eqref{eq:n-wave kinetic equation} is
consistent with thermalization.

The nonvanishing $\mu_k$ and $\gamma_k$ impose that only $p$-wave resonances allow for energy redistribution and lead to thermalization. Here resonances refer to the interaction process conserving energy and momentum, i.e.,
\begin{subequations}\label{eq:resonance conditions}
\begin{align}
  k_{1}\pm\cdots \pm k_{p}=0,\label{eq:on wavevector} \\
  \tilde{\omega}_{k_1}\pm\cdots\pm\tilde{\omega}_{k_p}=0. \label{eq:on frequency}
\end{align}
\end{subequations}
In generic many-body systems, resonances typically exist and are expected to group into a network including all dressed waves, leading to thermalization.  Employing the Eq.~\eqref{eq:n-wave kinetic equation} we conclude that the thermalization time scales as \cite{Wang2020,Wang2021}
\begin{equation}\label{eq:scaleing law}
  \tau_{\mathrm{th}}\propto \gamma_k^{-1}\propto g^{-2}\eta^3(\eta^2+1)^{p-2}.
\end{equation}

The result \eqref{eq:scaleing law} provides a more precise and inclusive conclusion, that is, resonances among dressed waves $\tilde{a}_k$ dominate the thermalization. In the weakly interacting regime (WIR), the renormalization factor $\eta\sim 1$, and therefore $\tau_{\mathrm{th}}\propto g^{-2}$. This result has been already confirmed in most recent works which conclude that resonances (among bare waves $a_k$) dominate thermalization \cite{Onorato2015,Lvov2018,Pistone2018,Fu2019,Pistone2019,Wang2020}. In the strongly interacting regime (SIR), we have $\eta\sim g^{1/p}$, thereby the thermalization time scales as $\tau_{\mathrm{th}}\propto g^{-1/p}$. In summary,
\begin{equation}\label{eq:summary}
  \tau_{\mathrm{th}}\propto
  \left\{
    \begin{array}{ll}
      g^{-2}\propto \lambda^{-2}\varepsilon^{2-p}, & \hbox{WIR;} \\
      g^{-1/p}\propto\lambda^{-1/p}\varepsilon^{1/p-1/2}, & \hbox{SIR.}
    \end{array}
  \right.
\end{equation}

\emph{Numerical verification}.---
In the rest of the paper, we test the previous ideas and analytical results in numerical experiments. For a FPUT-like system, its equivalent system tells that the thermal state is given by Eq.~\eqref{eq:202202131615}, from which we obtain $\tilde{\omega}_{k}\expval*{\tilde{a}_k\tilde{a}_l}=\theta\delta_{kl}$, namely, energy equipartition. To track the level of thermalization where the system arrives, we adopt the effective degrees of freedom proposed in Ref.~\cite{Benettin2011}, i.e., $d_{\mathrm{eff}}(t)=2\xi(t)e^{s(t)}/N$, where $s(t)=-\sum_{k=N/2}^{N}w_k(t)\ln w_k(t)$ is the spectral entropy, where $w_k=\bar{e}_k(t)/\sum_{l=N/2}^{N}\bar{e}_l(t)$, $\xi(t)=2\sum_{k=N/2}^{N}\bar{e}_{k}(t)/\sum_{k=1}^{N}\bar{e}_{k}(t)$ and $\bar{e}_k(t)=1/(T-\mu T)\int_{\mu T}^{T}e_k(t)\mathrm{d}t$. Here $e_k(t)$ could be the energy of either $a_k$ or $\tilde{a}_k$, and parameter $\mu$ controls the size of the time window for averaging and is fixed at $\mu=2/3$ in simulation. In the following section, we will draw a comparison between the effective degrees of freedom computed respectively from $a_k$ and $\tilde{a}_k$. The results are denoted correspondingly by $d_{\mathrm{eff}}(\vb*{a},t)$ and $d_{\mathrm{eff}}(\tilde{\vb*{a}},t)$. It is expected that $d_{\mathrm{eff}}(\tilde{\vb*{a}},t)$ when the system thermalizes.

In numerical simulations, the equations of motion are integrated by the eighth-order Yoshida algorithm \cite{Yoshida1990}. The time step is adapted with energy density to ensure that the absolute error in energy density is less than $10^{-4}$. The time step in our simulation ranges from $10^{-1}$ to $10^{-4}$. The initial energy is distributed evenly among $10\%$ waves of the lowest-frequency. To reduce the fluctuations, as done in Ref.~\cite{Benettin2011}, we average the observables on 128 random initial states. For the sake of convenience, we will fix $\lambda$ to be unit and investigate the dependence of $\tau_{\mathrm{th}}$ on $\varepsilon$ [See Eq.~\eqref{eq:summary}].

\begin{figure}[!t]
  \centering
  \includegraphics[width=0.95\columnwidth]{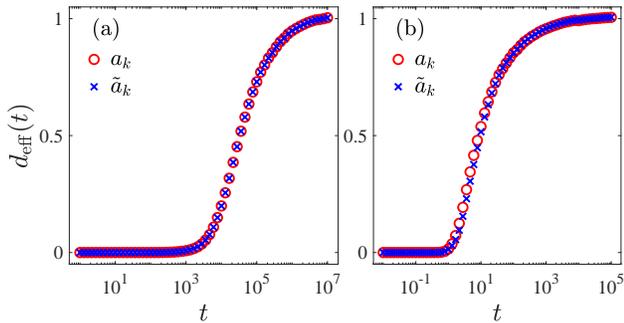}\\
  \caption{The effective degree of freedoms $d_{\mathrm{eff}}(t)$ versus time $t$ after exciting $10\%$ modes of the lowest frequency. Numerical experiments are conducted on the system $\mathscr{H}_{4}$ of $N=1024$, and energy densities are $\varepsilon=0.01$ (left) and $\varepsilon=100$ (right). The results colored red is computed from $\vb*{a}$ and these colored blue is computed from $\tilde{\vb*{a}}$}\label{fig:Figure-1}
\end{figure}

Before moving on, we first clarify a question: can waves $a_k$ be used to compute the level of thermalization? In the following, we address this question with a positive answer. On the one hand, the equipartition within dressed waves $\tilde{a}_k$ can always serve as an indicator of thermalization in the sense that $\tilde{\omega}_{k}\expval*{\tilde{a}_k^*\tilde{a}}_{\mathrm{GE}}\simeq \theta$. On the other hand, from the relation (\ref{eq:transformation}) and the GGE \eqref{eq:202202131615}, we obtain that the ratio $[\omega_k\expval*{a_k^*a_k}_{\mathrm{GE}}/[\tilde{\omega}_k\expval*{\tilde{a}_k^*\tilde{a}_k}_{\mathrm{GE}}]=(\eta^2+1)/(2\eta^2)$ is a constant, which implies that the level of thermalization will not change if one replaces $\tilde{\vb*{a}}$ with $\vb*{a}$. Numerical experiments are conducted on the system $\mathscr{H}_{4}$ with $N=1024$ to verify our prediction. Figure \ref{fig:Figure-1} shows the effective degree of freedoms $d_{\mathrm{eff}}(t)$. Two energy densities, i.e., $\varepsilon=0.01$ (left), $100$ (right), respectively selected from the WIR and SIR, are considered. What can be seen in Fig.~\ref{fig:Figure-1} is that the levels of thermalization computed from $\vb*{a}$ (red circle) and $\tilde{\vb*{a}}$ (blue cross) almost always coincide. This result confirms our prediction that $d_{\mathrm{eff}}(\tilde{\vb*{a}},t)$ and $d_{\mathrm{eff}}(\vb*{a},t)$ are equivalent, whereby affirms our hypothesis \eqref{eq:202202131615} indirectly. Further, we can conclude that prethermalization occurs in the WIR and SIR. In the remainder of this article, we will adopt the effective degree of freedoms computed from $\vb*{a}$.

\begin{figure}[!t]
  \centering
  \includegraphics[width=0.95\columnwidth]{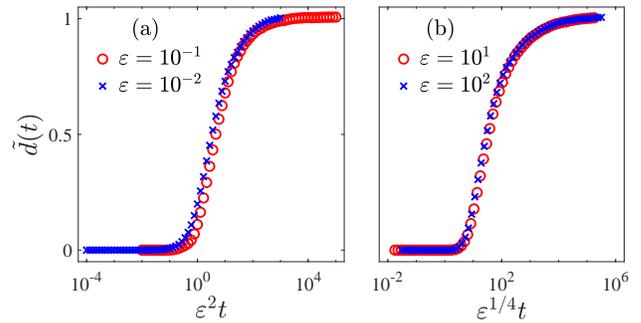}\\
  \caption{The effective degree of freedoms $d_{\mathrm{eff}}(t)$ as a function of (a) $\varepsilon^2t$ in weak-interaction regime, and (b) $\varepsilon^{1/4} t$ in strong-interaction regime. Numerical experiments are conducted on the system $\mathscr{H}_{4}$ with $N=1024$, and energy densities are listed in the plot.}\label{fig:Figure-2}
\end{figure}

Now it is time to consider the properties of $d_{\mathrm{eff}}(t)$. As previously, numerical experiments are conducted on the system $\mathscr{H}_{4}$ with $N=1024$. In Fig.~(\ref{fig:Figure-2}), we show the $d_{\mathrm{eff}}(t)$ versus time scaled by energy density (whose value is already listed in the figure) $\varepsilon^{\alpha}t$, where $\alpha=2$ in the WIR (left) and $\alpha=1/4$ in the SIR (right). As shown in the figure, all the functions $d_{\mathrm{eff}}(t)$ monotonically increase from $0$ to $1$. What stands out in this figure is that, in both regimes, there is good data collapse for different energy densities. These results imply that the thermalization time has different time scales in the WIR and SIR. It is reasonable to define the thermalization time as the moment when $d_{\mathrm{eff}}(t)$ first reaches a predefined threshold $d_{\mathrm{c}}$. The specific value of the threshold $d_{\mathrm{c}}$ does not affect the scales. Nevertheless, the threshold value $d_{\mathrm{c}}$ should be large enough since thermalization will reach a considerable level in a trivial amount of time in the SIR. In our computation, we set the threshold $d_{\mathrm{c}}$ to be $0.9$.

\begin{figure}[!t]
  \centering
  \includegraphics[width=0.95\columnwidth]{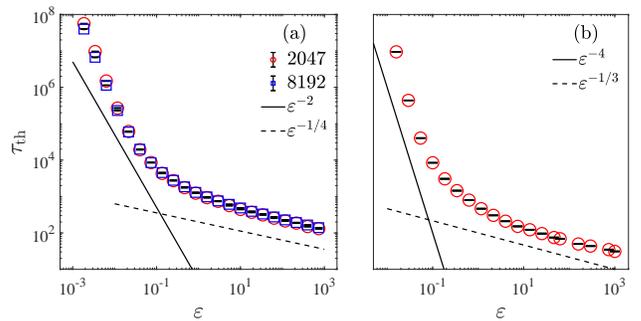}
  \caption{The thermalization time $\tau_{\mathrm{th}}$ as a function of $\varepsilon$ in log-log scale for system $\mathscr{H}_{4}$ with $N=2048$ (circles), $8192$ (squares), and (b) systems $\mathscr{H}_{6}$ with $N=2048$. The straight lines correspond to prediction obtained with Eq.~\eqref{eq:summary}.}\label{fig:Figure-3}
\end{figure}

At last, we present in Fig.~(\ref{fig:Figure-3}) the thermalization time $\tau_{\mathrm{th}}$ as a function of energy density $\varepsilon$ for systems $\mathscr{H}_{4}$ (left) and $\mathscr{H}_{6}$ (right). The theoretical predictions for the WIR (SIR) are plotted by solid (dashed) lines. It can be seen from Fig.~\ref{fig:Figure-3}(a) that prediction agrees well with the data, i.e., $\tau_{\mathrm{th}}\propto\varepsilon^{-2}$ in the WIR and $T_{\mathrm{eq}}\propto \varepsilon^{-1/4}$ in the SIR. Meanwhile, in the WIR, some deviations from the prediction are observed. Many works have reported this deviation and attributed it to the finite-size effect \cite{Wang2020,Fu2019}.  To illustrate this, the computation with a larger size, i.e., $N=8192$, is performed. The larger the system size is, the better the prediction $\tau_{\mathrm{th}}\propto\varepsilon^{-2}$ agrees with the data. Note that no finite-size effect exists under strong interactions since the broadening of frequencies significantly suppresses it. In Fig.~\ref{fig:Figure-3}(b), we report the result of the system $\mathscr{H}_{6}$, which is also consistent with the prediction, i.e., $\tau_{\mathrm{th}}\propto\varepsilon^{-4}$ in the WIR and $T_{\mathrm{eq}}\propto \varepsilon^{-1/3}$ in the SIR.

The double scaling of the thermalization time has already been observed in systems $\mathscr{H}_{4}$ with small size \cite{Lvov2018}. The authors argued that the time scale crosses over from $\tau_{\mathrm{th}}\propto \varepsilon^{-4}$  to $\tau_{\mathrm{th}}\propto\varepsilon^{-1}$ as the interaction increases. For small size, six-wave resonances, instead of four-wave ones, dominate the thermalization and lead to the time scale $\tau_{\mathrm{th}}\propto\varepsilon^{-4}$. However, energy density at which $\tau_{\mathrm{th}}\propto\varepsilon^{-1}$ in Ref.~\cite{Lvov2018} is $\sim 1$, which is far from the SIR. It is expected the thermalization time will eventually scale as $\tau_{\mathrm{th}}\propto\varepsilon^{-1/4}$ when the interaction becomes stronger.

\emph{Conclusion and discussion}.---In summary, we put forward a general method dealing with the non-equilibrium dynamics in a system of interacting waves. The idea lies in mapping an interacting system into a statistically equivalent one but with weaker interaction. The map provides an alternative path to identify dynamics of observables of the original system. After doing so, we find that prethermalization occurs not only in the WIR but also in the SIR and that resonances among dressed waves dominate the thermalization. Applying the second-order perturbation theory to the statistically equivalent system, we obtain the Zakharov equation describing the irreversible process towards equilibrium, from which we observe a different scaling for the thermalization time in the SIR.\par
Our method may provide an alternative way to probe many-body localization in disordered many-body systems. On the one hand, the trivial interactions dress the Anderson localized modes, enhancing the disorder. On the other hand, they also statistically weaken the interactions. These two effects may lead to stable localized quasi-integrals of motion. Finally, we would like to stress that our method can translate directly into the quantum version.

We acknowledge support by NSFC (Grants No. $\mathfrak{11975190}$ and No. $\mathfrak{11975189}$).
\bibliography{reference}

\end{document}